\documentclass[hyper]{JHEP}
\usepackage{graphicx,amssymb,bm,latexsym,amsmath}
\usepackage{epstopdf}
\usepackage{caption}
\usepackage{subcaption}
\usepackage[toc,page]{appendix}
\newcommand{\bej}[1]{ \begin{equation}\label{#1} }
\newcommand{\eej}{\end{equation}}
\newcommand{\beaj}[1]{\begin{eqnarray}\label{#1} }
\newcommand{\eeaj}{\end{eqnarray}}
\newcommand{\eq}[1]{(\ref{#1})}
\def\ZZZ{{\hskip-3pt\hbox{ Z\kern-1.6mm Z}}}
\def\zzz{{\hskip-3pt\hbox{ z\kern-1mm z}}}

\newcommand{\bd}{\bar{\rm D}}

\newcommand{\N}{\frac{m_{2}}{k_{2}}-\frac{m_{1}}{k_{1}}}

\newcommand{\be}{\begin{equation}}
\newcommand{\ee}{\end{equation}}
\newcommand{\ben}{\begin{eqnarray}\displaystyle}
\newcommand{\een}{\end{eqnarray}}

\def\one{{\hbox{ 1\kern-.8mm l}}}
\def\zero{{\hbox{ 0\kern-1.5mm 0}}}

\def\be{\begin{equation}}       
\def\ee{\end{equation}}         

\def\bea{\begin{eqnarray}}      
\def\eea{\end{eqnarray}}
                                
\def\ba{\begin{array}}
\def\ea{\end{array}}
\def\bd{\begin{displaymath}}
\def\ed{\end{displaymath}}

\def\eq{\begin{equation}}
\def\eqe{\end{equation}}
\def\eqa{\begin{eqnarray}}
\def\eqae{\end{eqnarray}}
\def\ena{\end{eqnarray}}

\def\Tr{{\rm Tr}}

\def\unit{1 \hskip-.3em \raise2pt\hbox{$ \scriptstyle |$ } }
\def\bd{\begin{displaymath}}
\def\ed{\end{displaymath}}

\def\6{\partial}
\def\N4{{\cal N}=4}

\def\bop#1{\setbox0=\hbox{$#1M$}\mkern1.5mu
        \vbox{\hrule height0pt depth.04\ht0
        \hbox{\vrule width.04\ht0 height.9\ht0 \kern.9\ht0
        \vrule width.04\ht0}\hrule height.04\ht0}\mkern1.5mu}
\def\>{\rangle} %right angle

\def\<{\langle} %left angle
\def\Dsl{D \hskip-.6em \raise1pt\hbox{$ / $ } }
\def\to{\rightarrow}

\def\+{\oplus}

\def\Tr{{\rm Tr}\, }

\def\as2{AdS_3\times S^3_1 \times S^3_2}

\title{On circular strings in $(AdS_3 \times S^3)_{\varkappa}$ }
\author{Aritra Banerjee$^1$,  Kamal L.  Panigrahi$^{1,2}$\\
$^1$Department of Physics,\\Indian Institute of Technology Kharagpur,\\
Kharagpur-721 302, India\\
$^2$Theory Group-DESY  Hamburg\\
Notkestrasse 85, D-22603, Hamburg\\
Germany\\
Email: \email{aritra, panigrahi@phy.iitkgp.ernet.in}}

\abstract{The so called one-parameter (often called $\varkappa$) deformed $AdS$ string sigma models have attracted a lot of attention lately in the study of integrability in string theory.  We construct various circular string solutions in the $(AdS_3 \times S^3)_{\varkappa}$ background and describe the characteristics of such solutions qualitatively. We study the Bohr-Sommerfeld like quantization for these string states to characterise the motion. 
Further we find a `long' string limit of such circular strings in the $\varkappa$-deformed $AdS_3$ and find a novel dependence of the oscillation number 
on the energy in the next to leading order expansion.}


\begin{document}
\section{Introduction}
The study of string spectrum on semisymmetric superspaces \cite{Zarembo:2010sg} have benefitted greatly from the study of integrability. Integrable structures have been studied in detail on both sides of the AdS/CFT correspondence over the years and remains one of the most active areas of research in string theory.  The most studied example of the AdS/CFT correspondence remains the one between spectrum of closed superstrings 
(supergravity) in $AdS_5\times S^5$ background and gauge invariant operators in four dimensional $\mathcal{N}= 4$ Supersymmetric Yang-Mills (SYM) theory \cite{Maldacena:1997re} based on the gauge group $SU(N)$. In the string side, the string dynamics in $AdS_5\times S^5$ can be described by a supercoset sigma model and the integrability of that has been investigated  \cite{Bena:2003wd}  in detail. Similarly the integrability of the dual SYM theory has also been explored extensively. Of course in the traditional sense of AdS/CFT, relating all the string states to the dual gauge theory operators appears to be a very tough job precisely because there are infinite tower of states in the string theory side. 
A probable way out is that in the large angular momentum or large R-charge limit both sides of the duality become more tractable. One of the perks of
this limit is that the anomalous dimension of operators in the SYM theory can be related to the dispersion relation between conserved charges of spinnings strings in the large charge limit.  On the other hand the dual gauge theory itself can be mapped to a particular type of integrable spin-chain system \cite{Minahan:2002ve}. So, using the gauge/gravity duality in general would allow to map rigidly rotating strings in gravity side to spin chain excitations in the gauge side \cite{Beisert:2003yb,Kazakov:2004qf,Zarembo:2004hp,
Beisert:2004hm,Arutyunov:2004vx,
Staudacher:2004tk,Beisert:2005fw,Beisert:2005tm}. This fact has generated a lot of interest in study of spinning string solutions in AdS and AdS-like exact string backgrounds.

Many types of classical string solutions have been studied in this context, lot of it arising from $AdS_5\times S^5$ string sigma models, and the dual spin chain excitations have been mapped. This includes well studied Giant Magnon \cite{Hofman:2006xt}, Folded Strings  \cite{Gubser:2002tv} and Spiky Strings \cite{Kruczenski:2004wg}  solutions. On the other side the circular pulsating string solutions \cite{Minahan:2002rc}  have been less worked out in this regard. These string solutions are time dependent ones as opposed to the other ones mentioned before. These solutions are dual to highly excited states in the dual spin chain picture. General circular pulsating strings in $S^5$ has been shown to correspond to states built out of the complex scalars in the $SO(6)$ sector of the dual SYM theory. For example a generic dual operator is made out of the chiral scalars $\Phi_1$,$\Phi_2$,$\Phi_3$ and has the form $\Tr (\Phi_1^{J_1}\Phi_2^{J_2}\Phi_3^{J_3} )$, where the $J_i$'s are the R-charges from the SYM theory (corresponding to the angular momenta along isometry directions of the sphere  for the string configuration).  Circular strings in AdS were introduced in \cite{Minahan:2002rc}, and generalized in  \cite{Khan:2003sm,Engquist:2003rn,
Arutyunov:2003za,Dimov:2004xi,Smedback:1998yn,Kruczenski:2004cn}. There has been a lot of generalisations of these solutions to various exact 
string backgrounds having different amounts of supersymmetry preserved, see for example \cite{Chen:2008qq,
Bobev:2004id,Arnaudov:2010by, Park:2005kt, {Pradhan:2013sja}}. Some one loop correction for such strings have been computed in detail also 
\cite{Beccaria:2010zn}. Also these solutions were discussed at length for the presence of a WZW term in the  string sigma model, in the context of string solutions in $AdS_3\times S^3$ with mixed three-form fluxes \cite{Banerjee:2014gga,Banerjee:2015bia}. We are concerned here about such circular strings in a one parameter deformation of  $AdS_5\times S^5$  background, the sigma model associated to which retains integrability of the original model. Such string solutions were preliminarily studied in \cite{Panigrahi:2014sia}. 

This novel one parameter deformed sigma model was first constructed in \cite{Delduc:2013qra}, reviving a few earlier proposals of a class of 
models put forward by Klimcik \cite{Klimcik:2002zj, Klimcik:2008eq, Klimcik:2014bta}. This particular model is completely different from integrable
deformations of AdS string sigma models that have been discussed before, including the TsT deformed scenarios and orbifolds 
\cite{Lunin:2005jy,Frolov:2005ty,Frolov:2005dj,Alday:2005ww}. In this case however, the deformation works by deforming the lie algebra itself 
by a continuous 
parameter, which is often referred to as a $q$-deformation. This replaces the symmetry algebra of the classical charges by its $q$-deformed version, 
which is then incorporated in to the superstring action for  $AdS_5\times S^5$  having a real deformation parameter. The currents associated with 
the deformed symmetry group element is modified by a linear operator $R$, which can be seen to obey the modified classical Yang-Baxter equation
(mCYBE). The amazing fact here is the classical integrability of the original model is preserved even in the deformed one. One can then read off the 
metric and NS-NS fields associated with this deformed structure. This background is often termed $\varkappa$ deformed  $AdS_5\times S^5$ , where 
$\varkappa \in [0,\infty)$ is a parameter related to $q$. The limit $\varkappa \to 0$ does give back the undeformed background, but in general the
deformed background is highly different from the undeformed one in many aspects. At the bosonic level, it breaks the original symmetry group
$SO(2,4)\times SO(6)$ to its Cartan subgroup $[U(1)]^3\times [U(1)]^3$. Various consistent truncations of the full background have been discussed 
at length in \cite{Hoare:2014pna}. There has been proposals of consistent type IIB supergravity solutions for $AdS_2\times S^2$ and $AdS_3\times S^3$ truncations \cite{Lunin:2014tsa}. But recently it was reported that instead of the usual supergravity equations of motion, the background satisfies
a particular deformed set of IIB equations \cite{Arutyunov:2015mqj,Wulff:2016tju}. For various issues concerning the integrable structure of this 
background, look at \cite{Arutynov:2014ota,Arutyunov:2014cra,Delduc:2014kha,Hollowood:2014rla,Hoare:2014oua, Arutyunov:2014jfa, Klimcik:2015gba,Arutyunov:2015qva,Kameyama:2015ufa, Pachol:2015mfa, Hoare:2016ibq}\footnote{One can look at \cite{Borsato:2016hud} and references thereof for an extensive introduction to this aspect.}.

Given the integrable and astounding nature of the deformed background, various rigidly rotating and pulsating strings have been investigated 
in detail. In a subspace of the full background, the giant magnon and spike solutions along with associated finite size corrections have been found \cite{Arutynov:2014ota, Banerjee:2014bca,Khouchen:2014kaa,Ahn:2014aqa,Kameyama:2014vma}. Since the background suffers from the presence of a singularity surface in the AdS space, a new coordinate system to handle this has been developed \cite{Kameyama:2014vma}. The deformed Neumann-Roschatius systems for the spinning strings in this model has been discussed at length in \cite{Arutyunov:2014cda}. The folded GKP like solutions 
were also found in \cite{Kameyama:2014vma} and generalised to N-spike strings in \cite{Banerjee:2015nha}. In both of these cases it was 
shown that in the `long' string limit, where the strings touch the singularity surface, the expression for cusp anomalous dimension does not reduce 
to the one in the $\varkappa\to 0$ limit. Not only classical string solutions, various minimal surfaces and wilson loops in this background have been found in \cite{Kameyama:2014vma, Kameyama:2014via, Bai:2014pya}. Study of three point correlators of such string states have been carried out in \cite{Bozhilov:2015kya, Bozhilov:2016owo}.  Also, more recently quark-antiquark potential in this background have been computed \cite{Kameyama:2016yuv}.

In this brief note, we are interested in the circular string solutions in the deformed background by solving the relevant F string equations of motion. 
We find the exact solutions in terms of elliptic functions corresponding to the string configurations and discuss their characteristics. The Bohr-Sommerfeld like quantization for such string states have been introduced in many places. Here, we mainly focus on the solutions themselves, corresponding to strings moving in $(AdS_3\times S^3)_\varkappa$. We show that the solutions in the two different $S^2$ subspaces of the deformed three sphere have completely different string configurations, albeit related by the discrete symmetries of the background itself. 
We also comment on the effect of adding an angular momentum to the solution along one of the remaining $S^1$ directions. We then solve for
circular string configurations in $(AdS_3)_\varkappa$  and define the charges and the oscillation number $N = \oint p~dq$ to characterise its motion. We subsequently try to find a `long' string limit of such solution and show that oscillation number has a completely new expansion in terms of energy, 
in this limit and can't be reduced to the known one in the limit $\varkappa \to 0$, in tune with earlier findings. 

The rest of the paper is organised as follows. In section 2, we discuss circular string solutions on the two different one-parameter deformed $S^2$ obtained by consistent truncations of $\varkappa$ deformed $S^3$. We show that the the two solutions are related by the ${\mathbb Z}_2$ transformation, which was the symmetry by which the two $S^2$s were related. We look at the Bohr-Sommerfeld like quantization and study the motion qualitatively. Section-3 is devoted to the study of similar solution in the deformed $AdS_3$ part of the geometry. We study a long string limit 
of such strings and find  a completely new scaling for such long strings which does not reduce to the usual form of oscillation number for $AdS$ 
strings in terms of energy in the next to leading order, though the leading order term reduces to that of the usual $AdS$ long strings in the 
$\varkappa\rightarrow 0$ limit. In section-5, we concluding with few comments and some future directives.

\section{Circular strings in $\varkappa$-deformed sphere}
\subsection{String solutions on deformed two spheres}
We start with the total $\varkappa$ deformed $AdS_3\times S^3$ geometry
as outlined in  \cite{Hoare:2014pna}, which has the form of
\be 
ds^2 = -h(\rho) dt^2 + f(\rho)d\rho^2+\rho^2 d\psi^2 + \tilde{h}(r)d\varphi^2+\tilde{f}(r)dr^2+ r^2 d\phi^2 \ , \label{metric1}
\ee
with  
\be
h(\rho) = \frac{1+\rho^2}{1-\varkappa^2\rho^2},~~~f(\rho) = \frac{1}{(1+\rho^2)(1-\varkappa^2\rho^2)} \ .
\ee
The $\tilde{h}$ and $\tilde{f}$ can be found by analytically continuing $\rho \to i r$. The NS-NS fluxes associated to this solution does not survive in this consistent truncation of the total deformed $AdS_5 \times S^5$ background. Starting with a string moving only in $\mathbb{R}\times S^3$ ($\rho = 0$) by putting $r=\cos\theta$, we study particular circular string solutions in this deformed geometry. Now the metric on the deformed sphere becomes,
\be
ds_{S^3}^2 = -dt^2 + \frac{1}{1 + \varkappa^2 \cos^2\theta}~d\theta^2 + \frac{\sin^2\theta}{1 + \varkappa^2 \cos^2\theta} ~d\varphi^2 + \cos^2\theta ~d\phi^2.
\ee
We start with the following ansatz for the circular string configuration
\be 
t= C_0 \tau, ~~~ \theta= \theta (\tau),~~~\varphi= m \sigma,~~~\phi = \text{constant}
\ee
We use the usual Polyakov action for the F string,
\begin{equation}
S=-\frac{\sqrt{\hat{\lambda}}}{4\pi}\int d\sigma d\tau
[\sqrt{-\gamma}\gamma^{\alpha \beta}g_{MN}\partial_{\alpha} X^M
\partial_{\beta}X^N ] \ ,
\end{equation}
where  $\hat{\lambda} = 
\lambda (1 + {\varkappa}^2)$ is the modified 't Hooft coupling for the deformed model and
$\gamma^{\alpha \beta}$ is the worldsheet metric. 
Variation of the action with respect to
$X^M$ gives us the following equations of motion
\begin{eqnarray}
2\partial_{\alpha}(\eta^{\alpha \beta} \partial_{\beta}X^Ng_{KN})
&-& \eta^{\alpha \beta} \partial_{\alpha} X^M \partial_{\beta}
X^N\partial_K g_{MN} =0 \ ,
\end{eqnarray}
and variation with respect to the metric gives the two Virasoro
constraints,
\begin{eqnarray}
g_{MN}(\partial_{\tau}X^M \partial_{\tau}X^N +
\partial_{\sigma}X^M \partial_{\sigma}X^N)&=&0 \ ,\label{v1} \\ 
g_{MN}(\partial_{\tau}X^M \partial_{\sigma}X^N)&=&0 \label{v2}\ .
\end{eqnarray}
We use the conformal gauge (i.e.
$\sqrt{-\gamma}\gamma^{\alpha \beta}=\eta^{\alpha \beta}$) with
$\eta^{\tau \tau}=-1$, $\eta^{\sigma \sigma}=1$ and $\eta^{\tau
\sigma}=\eta^{\sigma \tau}=0$) to solve the equations of motion.
One can explicitly check from the string equations of motion that the above configuration is completely consistent with the virasoro constraints with particular choice of integration constants and indeed signifies a circular string moving in the $\mathbb{R}\times S^2_{\varphi}$ subspace of the total $(\mathbb{R}\times S^3)_\varkappa$.  The equations of motion from the sigma model for the above string leads to 
\be
\frac{\dot{\theta}^2}{1 + \varkappa^2 \cos^2\theta}+ \frac{m^2 \sin^2\theta}{1 + \varkappa^2 \cos^2\theta}  = \mathcal{E}^2
\ee
Here the $\mathcal{E}$ is the energy rescaled by the modified sigma model coupling constant $\sqrt{\hat{\lambda}}$. Remembering $\hat{T} = T\sqrt{1+\varkappa^2}$, it could be written in the form,
\be
\mathcal{E} = \frac{E}{\sqrt{\lambda}\sqrt{1+\varkappa^2}} = \frac{\mathcal{E}_0}{\sqrt{1+\varkappa^2}}.
\ee
Where $E$ is simply the conserved Noether charge associated to shift in $t$. Now this new scaling of charges adds further subtleties to the limits we would impose on the solutions here. We will discuss them at length at various parts of this note. Remember here that the $\varkappa = 0$ expression ($\mathcal{E}_0$) is usually called the `semiclassical' value of the conserved charge in the large `t Hooft coupling limit. But inclusion of the $\varkappa$ dependent factor in the coupling itself makes it an object to be explored cautiously. But as a starter we can see that $\mathcal{E}<\mathcal{E}_0$, since $\varkappa\in[0,\infty)$

To solve the above equation of motion, we substitute $x = \sin \theta$ and put the equation in the following form
\be
\dot{x}^2 = (m^2 + \mathcal{E}^2\varkappa^2)(x^2- R_+)(x^2 - R_-) \ ,
\ee
where the two roots are given by the following,
\be
R_- = \frac{\mathcal{E}^2(1+\varkappa^2)}{m^2+\mathcal{E}^2\varkappa^2},~~~~R_+ = 1.
\ee
With this, the solution of the string motion can be put in the form of a Jacobi Sn function
\footnote{In our notation, $\textbf{sn}(z|m)$ is the solution of the equation $P'(z)^2 = (1-P^2(z))(1-mP^2(z)) $}
, namely
\be
x_\varphi = \sin \theta_\varphi = \sqrt{R_-} ~\textbf{sn} \left(\sqrt{m^2+\mathcal{E}^2\varkappa^2}\sqrt{R_+}\tau | \frac{R_-}{R_+}\right)  \ . \label{sol1}
\ee
Now the condition on the solution to have an oscillating nature can be found from the real periodicity condition on the Jacobi function in the form
\be
0< \frac{\mathcal{E}^2(1+\varkappa^2)}{m^2+\mathcal{E}^2\varkappa^2} <1.
\ee
This leads obviously to the condition
\be
\mathcal{E}^2< m^2 
\ee
Which can be seen to reduce exactly to the condition of oscillating solution mentioned in \cite{Beccaria:2010zn} for $\varkappa = 0$ limit, i.e. $\mathcal{E}_0^2< m^2 $, with $m \in \mathbb{Z}$, which explicitly means that strings in the two-sphere can only have the `small' string limit. Here the presence of the $\varkappa$ adds special limits to this condition also. The rescaled energy is of course bound from above by the said inequality. But the inequality can be seen to be valid for all values of $\varkappa$ provided the original energy $\mathcal{E}_0$ is small. This will be crucial in the later discussions.

Now we move on to the other consistent truncation of the deformed $\mathbb{R}\times S^3$, which we will call $S^2_{\phi}$. The metric in this case is given by the following
\be
ds^2 = -dt^2 + \frac{1}{1 + \varkappa^2 \cos^2\theta}~d\theta^2 +  \cos^2\theta ~d\phi^2.
\ee
Taking an ansatz similar as before, 
\be
t= C_1 \tau, ~~~ \theta= \theta (\tau),~~~\phi = m\sigma \ ,
\ee
we arrive at the following equation of motion for $\theta$
\be
\frac{\dot{\theta}^2}{1 + \varkappa^2 \cos^2\theta}+ m^2 \cos^2\theta = \mathcal{E}^2 \ ,
\ee
where $\mathcal{E}$ is defined as before. This equation is little more involved than the last one. To solve this, we again put $\cos\theta = x$ and 
write the above equation in the form
\be
\dot{x}^2 = m^2\varkappa^2(x^2 -R_1)(x^2 -R_2)(x^2 -R_3).
\ee
Here we have the roots in the form,
\be 
R_1 = -\frac{1}{\varkappa^2},~~~R_2 = \frac{ \mathcal{E}^2}{m^2},~~~R_3=1.
\ee
Integrating the equation of motion is a little trickier, the direct integration leads to the expression,
\be
\tau = \frac{1}{\sqrt{m^2+\mathcal{E}^2\varkappa^2}}~\mathbb{F}\bigg[ \sin^{-1}\sqrt{\frac{(m^2+\mathcal{E}^2\varkappa^2)x^2}{\mathcal{E}^2(1+\varkappa^2 x^2)}}, \frac{(1+\varkappa^2)\mathcal{E}^2}{m^2+\mathcal{E}^2\varkappa^2}\bigg] \ .
\ee
After a little algebra, we can extract the follwing solution easily in the form of a Jacobi ${\bf sd}$
\footnote{In our notation, $\textbf{sd}(z|m)$ is the solution of $P''(z) +P(z)[2m(1-m)P^2(z)-2m+1]=0$}
 function
\be 
\cot \theta_\phi = \sqrt{\frac{\mathcal{E}^2}{m^2+\mathcal{E}^2\varkappa^2}}~\textbf{sd} \left(\sqrt{m^2+\mathcal{E}^2\varkappa^2}~\tau |~ \frac{\mathcal{E}^2(1+\varkappa^2)}{m^2+\mathcal{E}^2\varkappa^2} \right)  \ ,
\label{sol2}
\ee
where we have used the well known identity concerning Jacobi functions
\be
\textbf{dn}^2(u|~ \mathcal{X}^2)+ \mathcal{X}^2~\textbf{sn}^2(u|~ \mathcal{X}^2)= 1.
\ee
The other cause of writing the solution in this way will be clear as we go along. One can note that the condition for having an oscillatory solution remains the same as for the string moving in the other two sphere as we have discussed.

These string solutions can be shown to inherit the symmetries of the deformed background. As explained in the \cite{Hoare:2014pna} we can remember that the different two spheres $S^2_{\varphi}$ and $S^2_{\phi}$ are related to each other by the discrete $\mathbb{Z}_2$ transformations
\be
\varphi \to \phi~;~~\cos\theta\to \sqrt{\frac{\sin^2\theta}{1+\varkappa^2\cos^2\theta}}.
\ee
While the first one of these is actually trivial, one can explicitly check that the second transformation is completely valid for our circular string solutions (\ref{sol1}) and (\ref{sol2}), again implementing identities involving elliptic Jacobi functions. One might recall that in \cite{Banerjee:2014bca} the same mapping was shown to exist for giant magnon solutions on the two 2-spheres.

\subsection{Semiclassical quantization}
\begin{figure}

        \centering
        \begin{subfigure}[b]{0.5\textwidth}
                \includegraphics[width=\textwidth]{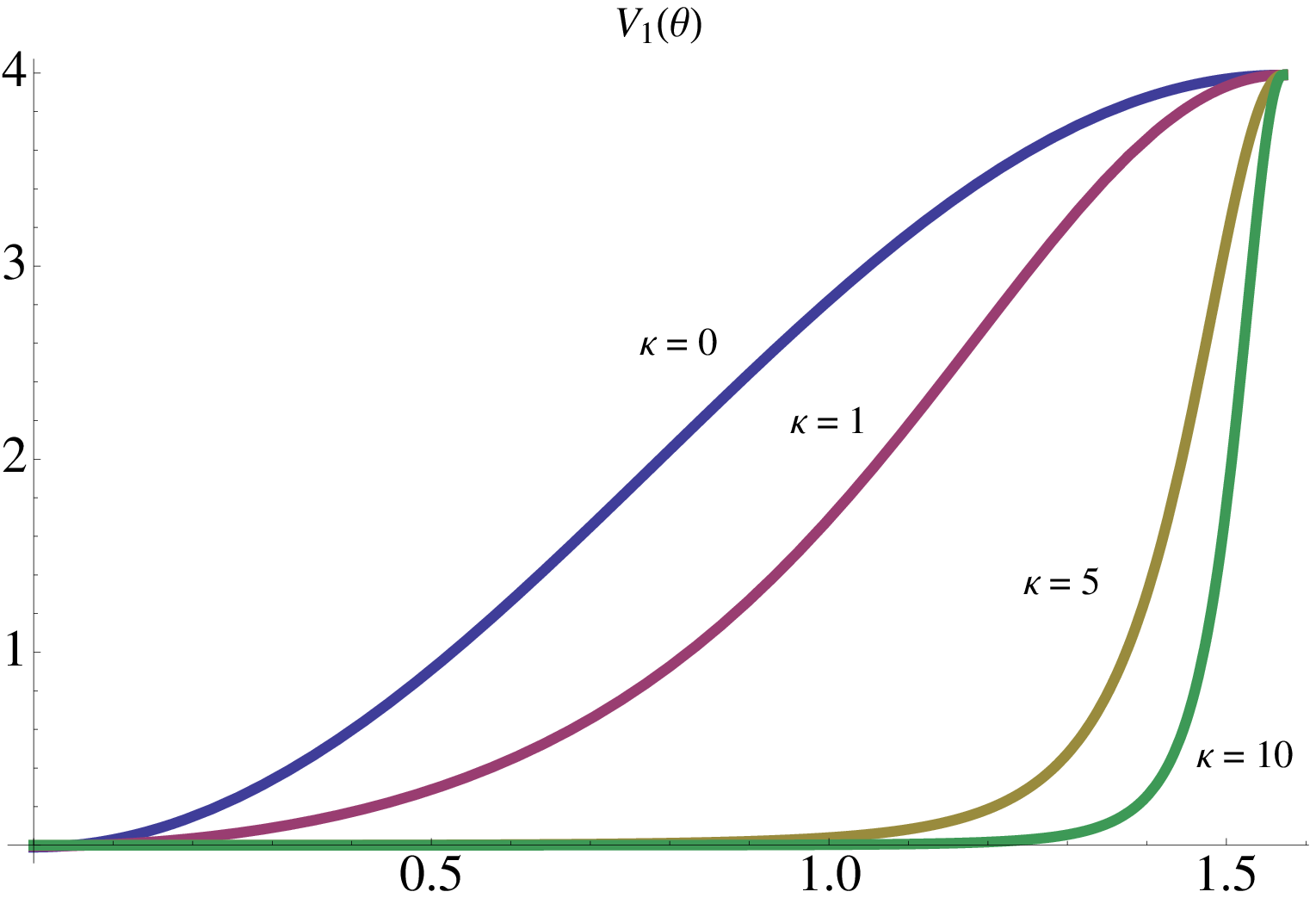}
                \caption{}
                \label{fig:1.1}
        \end{subfigure}%
        ~ %add desired spacing between images, e. g. ~, \quad, \qquad, \hfill etc.
        \begin{subfigure}[b]{0.5\textwidth}
                \includegraphics[width=\textwidth]{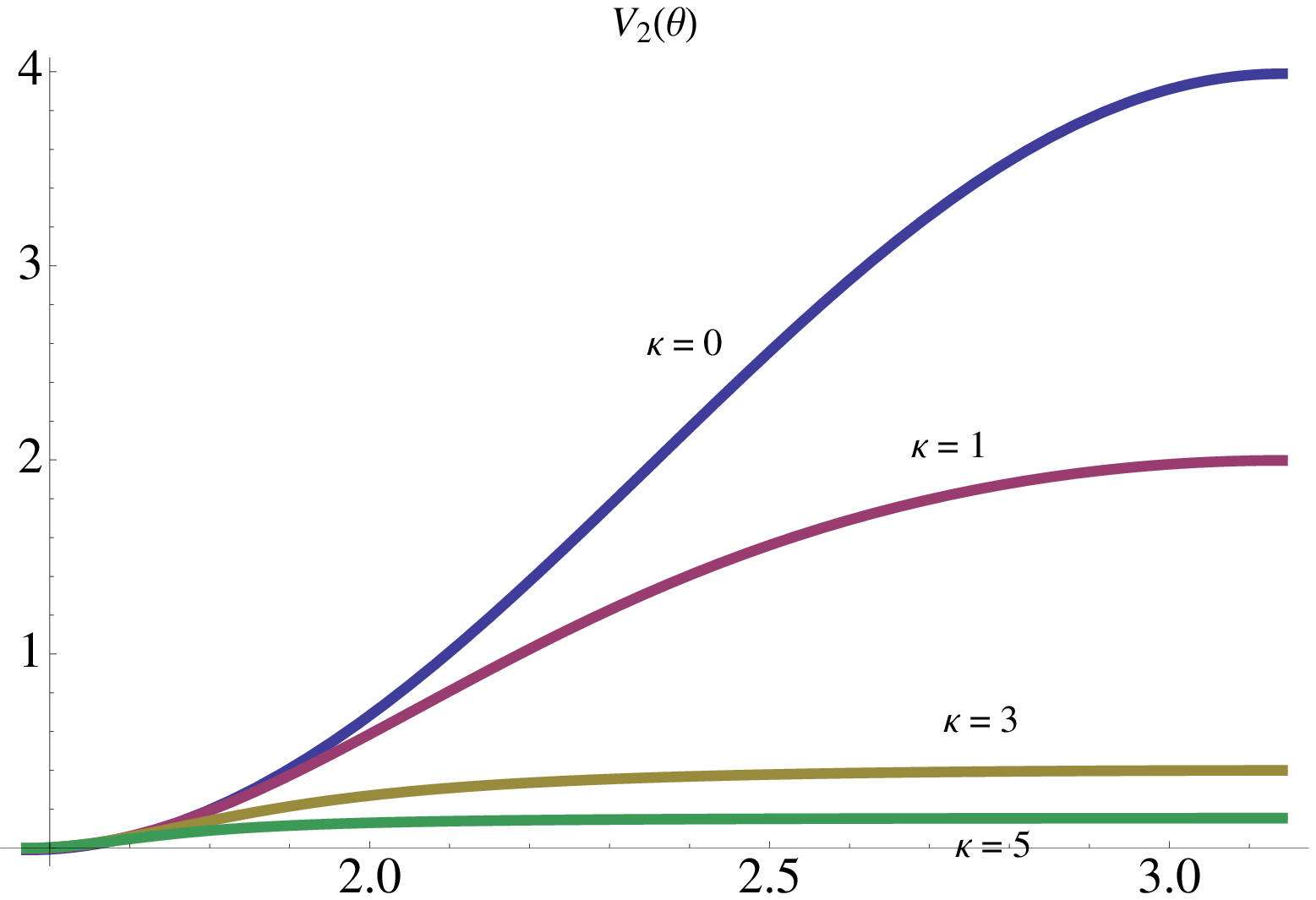}
                \caption{}
                \label{fig:1.2}
        \end{subfigure}

        \caption{The potentials corresponding to (\ref{v1}) and (\ref{v2}) are plotted here w.r.t. $\theta$ for different values of $\varkappa$. One can easily see that as for the first one,$V_\phi(\theta)$, becomes steeper as we increase the value of $\varkappa$. For the other case the potentials are more flattened as we increase $\varkappa$. For both the cases the maxima occurs at the same exact place as the undeformed potential. Both cases provide evidence that for larger $\varkappa$, the oscillatory solutions die out. For both of the figures $\mathcal{E} = 0.1$ and $m=2$.}\label{fig:1}
\end{figure}

For doing a sommerfeld like quantization of the strings discussed in the last section, we first start with the $s^2_{\varphi}$. Again, the adiabatic invariant associated to both the spheres is given simply by,
\be
\mathcal{N}= \frac{1}{2\pi}\oint~\Pi_\theta~d\theta \ ,
\ee
where we again define the semiclassical value of the adiabatic invariant scaled by the new coupling constant $\sqrt{\hat{\lambda}}$, but it can be shown to behave exactly like the original oscillation number (at $\varkappa =0$, which we can call $\mathcal{N}_0$) for finite values of $\varkappa$. 
It is crucial to remember that for consistent results, we can't let the value of $\varkappa$ to be very large, as we also know that for that limit the nature of the geometry changes altogether. Also $\Pi_\theta$ is the momentum associated to $\theta$. Now we can put ,
\be
\Pi_\theta= \frac{\dot{\theta}}{1+\varkappa^2\cos^2\theta}.
\ee
Now using this, the $\theta$ equation of motion can be written as 
\be
\Pi_\theta^2+ V_\varphi(\theta)  = 0.
\ee
This may be interpreted as an equation for a particle moving in a potential $V_{\varphi} (\theta)$ with ,
\be
V_\varphi(\theta) = -\frac{\mathcal{E}^2}{1+\varkappa^2\cos^2\theta}+\frac{m^2\sin^2\theta}{(1+\varkappa^2\cos^2\theta)^2}. \label{v1}
\ee
A closer look at the potential will reveal that it has a maximum at $\frac{\pi}{2}$ with a value ($m^2-\mathcal{E}^2$). So, at the level of the potential, there are two cases, one for which $\frac{\mathcal{E}^2}{m^2}<1$ so that there is a turning point and $\theta$ is limited to a maximum value. The other case is simply $\frac{\mathcal{E}^2}{m^2}>1$, where there is no turning point and the string oscillates all the way from equator to the pole of this deformed sphere. For our case, as we have demonstrated from the solution itself, the first case is physically acceptable. The turning point in this case simply lies at
\be
R_- = \frac{\mathcal{E}^2(1+\varkappa^2)}{m^2+\mathcal{E}^2\varkappa^2}.
\ee
So we can expect as $\varkappa$ increases the turning point also acquires a greater value, herebey making the potential steeper. As the $\varkappa\to \infty$, the value saturates around $1$, and as we remember the oscillatory behaviour of the solution is lost completely.
The potential has been plotted for completeness in  figure (\ref{fig:1}).

We can use the value of $\dot{\theta}$ from the relevant equation of motion and write the integral for the oscillation number using $\cos\theta = z$
\be
\mathcal{N}_\varphi =\frac{2}{\pi}\int \frac{dz}{1+\varkappa^2-\varkappa^2 z^2}\sqrt{\frac{\mathcal{E}^2(1+\varkappa^2-\varkappa^2 z^2)-m^2 z^2}{(1-z^2)}}.
\ee
Instead of directly dealing with the integral, one tries to find the derivative w.r.t. $m$ and evaluate the integral. With a little manipulation, we can write the derivative in the form
 \begin{equation}
\frac{\partial \mathcal{N}_\varphi}{\partial m} = \frac{2m }{\pi \varkappa^2 \sqrt{m^2 +\mathcal{E}^2\varkappa^2}}\Bigg[ \mathbf{K}\bigg( \frac{\mathcal{E}^2(1+\varkappa^2)}{\mathcal{E}^2\varkappa^2 + m^2} \bigg) -\mathbf{\Pi}\bigg( \frac{\mathcal{E}^2\varkappa^2}{\mathcal{E}^2\varkappa^2 + m^2}, 
\frac{(1+\varkappa^2)\mathcal{E}^2 }{\mathcal{E}^2\varkappa^2 + m^2}  \bigg)  \Bigg]
\end{equation}

Now we know that the condition on energy from the solution is $\mathcal{E}^2 < m^2$, that means we have to take small energy limit of the string. But since $\varkappa \in [0,\infty)$, the order of limits is very subtle here. To explain, let us start from the simplest possible limit where $\mathcal{E} \to 0$, but $\varkappa$ is finite, so that $\mathcal{E}\varkappa$ is small. We put $\mathcal{E}\varkappa = \alpha$ and write the above expression in this limit as

\be
\frac{\partial \mathcal{N}_\varphi}{\partial m} = \frac{2m}{\pi \varkappa^2 \sqrt{m^2 +\alpha^2}}\Bigg[ \mathbf{K}\bigg( \frac{\alpha^2}{\alpha^2 + m^2} \bigg) -\mathbf{\Pi}\bigg( \frac{\alpha^2}{\alpha^2 + m^2} ,\frac{\alpha^2}{\alpha^2 + m^2}     \bigg)\Bigg]
\ee
Now the problem boils down to taking $\alpha \to 0$ limit for the above expression.
We recall the series expressions of the elliptic integrals around arguments zero,
\be
\mathbf{K}(z) = \frac{\pi}{2}\sum_{n=0}^{\infty}\Bigg[\frac{(2n)!}{2^{2n}(n!)^2}    \Bigg]^2 z^n \ ,
 \ee
 
 \be
 \mathbf{\Pi}(z, z) =  \frac{\pi}{2}\sum_{n=0}^{\infty}\Bigg[\frac{(2n+1)((2n)!)^2}{4^{2n}(n!)^4} \Bigg] z^n \ .
 \ee
 Using these expansions and integrating over $m$, we can write simply,
 \be
 \mathcal{N}_\varphi = \frac{1}{\varkappa^2}\sqrt{\alpha^2 + m^2}\sum_{n=0}^{\infty} A_n \bigg( \frac{\alpha^2}{\alpha^2 + m^2} \bigg)^n \ ,
\ee
where 
\be
A_n = \frac{2^{1-4 n} n ((2 n)!)^2}{(2 n-1) (n!)^4} \ .
\ee
 But one could clearly see that since we take $\alpha \to 0$
and of course we have $A_n \to 0$ as $n \to \infty$, the series is bound to converge automatically, although it generates no additional constraints.

The other limit that can be taken is  $\mathcal{E}\to 0$ and $\varkappa \to \infty$, but $\alpha$ is finite. In this case the asymptotic expressions of elliptic integrals won't work, so we move on to the other extreme case. It is simply $\mathcal{E}\to 0$ and $\varkappa \to \infty$, but $\alpha \to \infty$. In this limit $\frac{\alpha^2}{\alpha^2 + m^2} \to 1$. Here the expression of $\frac{\partial \mathcal{N}_\varphi}{\partial m} $ becomes indeterminate since the prefactor is $0$, while the elliptic
integrals run to complex infinity. This is consistent with our previous observation that for large $\varkappa$ our solutions actually do not oscillate, i.e we can physically discard this limit. We will then stick to the first case as we have talked about earlier.

We can expand the expression for $\frac{\partial \mathcal{N}_\varphi}{\partial m} $  for small $\mathcal{E}$ and integrate over $m$ to find the expression of the oscillation number $\mathcal{N}_\varphi$. This expression can then be inverted to find the energy of the string in terms of 
\be 
\mathcal{E}  = \sqrt{2 m \mathcal{N}_\varphi}\left[ 1 + \frac{(-1+\varkappa^2)\mathcal{N}_\varphi}{8m} - \frac{(5+6\varkappa^2+5\varkappa^4)\mathcal{N}_\varphi^2}{128 m^2}  + \mathcal{O}(\mathcal{N}_\varphi^3)  \right]
\ee

Let us now move to the string moving in the other sphere, the $S^2_\phi$.  In this case also the $\theta$ equation of motion can be written in the form
\be
\Pi_\theta^2+ V_\phi(\theta)  = 0.
\ee
In this case the effective potential has a form
\be 
V_\phi(\theta)  = \frac{ m^2 \cos^2\theta-\mathcal{E}^2}{1+\varkappa^2\cos^2\theta}. \label{v2}
\ee
Which is, as expected, completely different that the $V_\varphi(\theta)$. This has a maximum value at $\theta = 0$ which is simply $\frac{m^2-\mathcal{E}^2}{1+\varkappa^2}$. Again for this case also the presence of the condition $\frac{\mathcal{E}^2}{m^2}<1$ ensures that there is one turning point for the potential. It can be seen from the figure (\ref{fig:1}) that as we increase the value of $\varkappa$ here, the maximum of the potential is reduced increasingly, and at $\varkappa\to \infty$ the oscillatory motion is simply stopped.
 
 In this case, the oscillation number can be given by
\be 
\mathcal{N}_\phi = \frac{1}{2\pi}\int \sqrt{\frac{\mathcal{E}^2- m^2 \cos^2\theta}{1+\varkappa^2\cos^2\theta}}~d\theta.
\ee
Following the procedure as outlined earlier, we could write again,
\be
\frac{\partial \mathcal{N}_\phi}{\partial m} = \frac{-2}{\pi \sqrt{m^2 +\mathcal{E}^2 \varkappa^2}}\Bigg[ m \mathbf{K}\bigg(\frac{\mathcal{E}^2(\varkappa^2+1)}{m^2 +\mathcal{E}^2 \varkappa^2}  \bigg)  + \frac{\mathcal{E}^2 - m^2}{m }\mathbf{\Pi} \bigg( \frac{\mathcal{E}^2}{m^2}, \frac{\mathcal{E}^2(\varkappa^2+1)}{m^2 +\mathcal{E}^2 \varkappa^2}  \bigg)\Bigg]
\ee
As in the last case the only plausible situation for expansion is the case where $\mathcal{E}$ is small and $\varkappa$ is finite. In this approximation we could expand the above expression and integrate over $m$ to find 
\be
\mathcal{N}_\phi =  \frac{\mathcal{E}^2}{2m}-\frac{\mathcal{E}^4(-1+\varkappa^2)}{8m^3}+\frac{\mathcal{E}^6(3-2\varkappa^2+3\varkappa^4)}{128m^3}+\mathcal{O}(\mathcal{E}^8) \ ,
\ee
which leads to the same expression of energy as in the above case. This is expected for the problem as however the two spheres are not equivalent, both of them reduce to the usual two sphere when we put the deformation parameter to be zero. As in both the cases we can put $\varkappa = 0$ and the expression matches with the one discussed in \cite{Beccaria:2010zn}.

\subsection{Circular strings in $(\mathbb{R}\times S^3)_\varkappa$: Adding angular momentum}
We now start discussing about circular strings propagating in the total three sphere, which would be more interesting. One of the peculiar phenomenon associated is that now the direction of the extra angular momenta will matter since the two 2-spheres are not equivalent to each other, but of course they could be related by discrete symmetries of the spacetime. For this reason these solutions demand some closer inspection. 

Here, we will only discuss the ansatz for the case where the string pulsates in $S^2_\varphi$ with the  extra angular momentum along $\phi$. The ansatz in this case will have the form,
\be
t = c_0\tau,~~\theta = \theta(\tau),~~\varphi = m\sigma,~~\phi= \omega \tau.
\ee
The equations of motion for $\theta$ has the following form
\be
\dot\theta^2  = (1+\varkappa^2 \cos^2\theta)(\mathcal{E}^2 - \frac{\mathcal{J}^2}{\cos^2\theta}) - m^2 \sin^2\theta \ ,
\ee 
where the conserved charge $\mathcal{E}$ corresponds to shift in $t$ and the angular momentum $\mathcal{J}$ is defined as
\be
\mathcal{J}  = \cos^2\theta ~\dot{\phi}
\ee
The equation for $\theta$ again looks like that of a particle in a potential oscillating between $(1+\varkappa^2)(\mathcal{E}^2 - \mathcal{J}^2)$ at $\theta = 0$ to $0$ at $\theta = \pi/2$.  With the substitution of $\sin\theta = x$ we can rewrite the equation of motion in the form 
\be
\dot{x}^2 = (m^2 + \mathcal{E}^2\varkappa^2)(x^2- \mathcal{R}_+)(x^2 - \mathcal{R}_-)
\ee
Where the roots in the above equation can be given by,
\be
\mathcal{R}_\pm = \frac{m^2+ \mathcal{E}^2 -\mathcal{J}^2\varkappa^2 + 2\mathcal{E}^2\varkappa^2 \pm \sqrt{(m^2-\mathcal{E}^2)^2+\mathcal{J}^4\varkappa^4+2\mathcal{J}^2(2m^2+(m^2+\mathcal{E}^2)\varkappa^2)}}{2(m^2+\mathcal{E}^2\varkappa^2)}
\ee
To solve the above equation, we demand the boundary condition $x(0)= 0$ and the final solution can be written in the form,
\be
\sin\theta(\tau)= \sqrt{\mathcal{R}_-}~\mathbf{sn}\left( \sqrt{m^2+\mathcal{E}^2\varkappa^2}\sqrt{\mathcal{R}_+}\tau~| \frac{\mathcal{R}_-}{\mathcal{R}_+}  \right).
\ee
 The condition for this to be a proper oscillating and periodic string solution reads
\be
0 <  \frac{\mathcal{R}_-}{\mathcal{R}_+}  <1
\ee

This boils down to the algebraic condition 
\be
(m^2-\mathcal{E}^2)^2+\mathcal{J}^4\varkappa^4+2\mathcal{J}^2(2m^2+(m^2+\mathcal{E}^2)\varkappa^2)>0
\ee

\begin{figure}

        \centering
        \begin{subfigure}[b]{0.7\textwidth}
                \includegraphics[width=\textwidth]{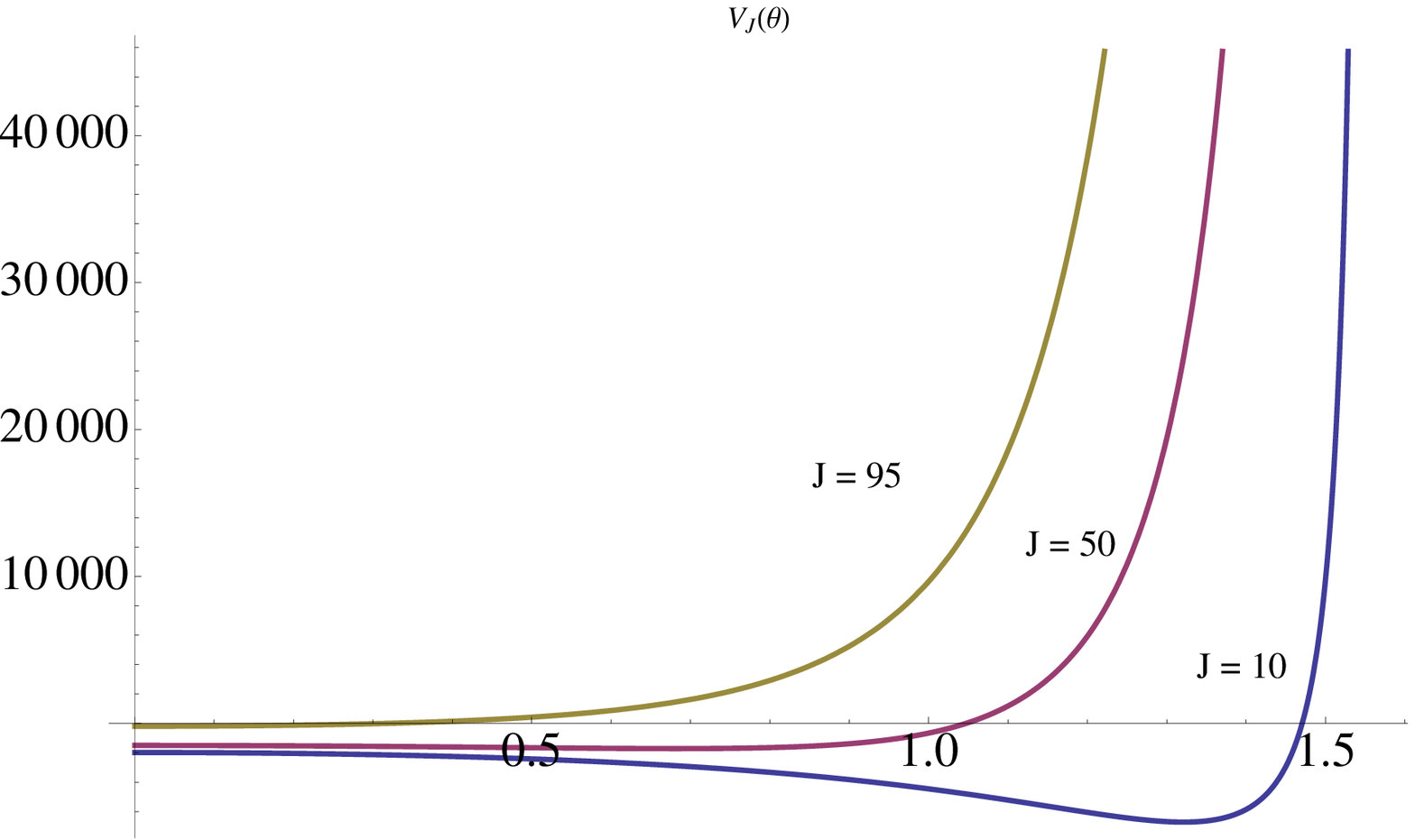}
                \caption{}
                \label{}
        \end{subfigure}

        \caption{The potentials corresponding to (\ref{potJ})  is plotted here w.r.t. $\theta$ for different values of $\mathcal{J}$. The values of other parameteres $(\mathcal{E},m,\varkappa) = (100,2,2) $ here. Notice that the potential blows up at $\frac{\pi}{2}$. }\label{fig:3}
\end{figure}

The above condition simply puts a bound on the conserved charges on the solution.
Also there is another condition that makes the roots $\mathcal{R}_\pm$ positive and it has a simple expression
\be
(1+\varkappa^2)(\mathcal{E}^2-\mathcal{J}^2)(m^2+\mathcal{E}^2\varkappa^2)>0
\ee
It can be seen that the original condition on the pulsating string $\mathcal{E}^2\geq\mathcal{J}^2$ stays when we put $\varkappa = 0$. 
This puts an upper bound on the value of angular momentum that can be switched on in this case. We can also integrate to get a closed form 
for the $\phi$ to concentrate on the dynamics with the angular momentum. We can write using the two above equations that
\be 
\frac{d\phi}{d\theta} = \frac{\mathcal{J}}{\cos\theta\sqrt{m^2+\mathcal{E}^2\varkappa^2}\sqrt{(\sin^2\theta - \mathcal{R}_+)(\sin^2\theta - \mathcal{R}_-)}} \ .
\ee
This can be integrated to find the expression for $\phi$ in terms of standard elliptic integral forms,
\be
\phi = - \frac{\mathcal{J}}{\sqrt{m^2+\mathcal{E}^2\varkappa^2}\sqrt{\mathcal{R}_+}}~\mathbf{\Pi}\left( \mathcal{R}_-, \sin^{-1}\left( \frac{1}{\sqrt{\mathcal{R}_-}} \sin\theta(\tau) \right)  ,\frac{\mathcal{R}_-}{\mathcal{R}_+}   \right)
\ee

Now we can see that there is only one adiabatic invariant available here expect for the extra angular momentum, which is the action variable associated to $\theta$. We can write that in the following way
\be
\mathcal{I}_\theta= \frac{\sqrt{\hat{\lambda}}}{2\pi}\oint~d\theta \sqrt{(1+\varkappa^2 \cos^2\theta)(\mathcal{E}^2 - \frac{\mathcal{J}^2}{\cos^2\theta}) - m^2 \sin^2\theta}
\ee
To make contact with string theory, we are interested in the large energy region of the solution and find out the corrections to the classical energy.  But with the new results, the calculations for expansion of charges appears very tricky as another paramter $\mathcal{J}$ has appeared in the equations of motion and the order of limits issue has to be taken care of. For the sake of completeness we can write the relevant effective potential in this case,
\be
\mathcal{V}(\theta)  = -\frac{(\mathcal{E}^2 - \frac{\mathcal{J}^2}{\cos^2\theta})}{(1+\varkappa^2 \cos^2\theta)} + \frac{m^2 \sin^2\theta}{(1+\varkappa^2 \cos^2\theta)^2}.\label{potJ}
\ee
The unique property of the potential is that it straightaway diverges at $\theta = \frac{\pi}{2}$. We have plotted the potential for different values of $\mathcal{J}$ in figure (\ref{fig:3}). One can easily see that the particle here oscillates from $-\frac{\mathcal{E}^2 - \mathcal{J}^2}{1+\varkappa^2}$ at $\theta = 0$ to $\infty$ at $\theta = \frac{\pi}{2}$. Since there is the constraint $\mathcal{E}^2 \geq \mathcal{J}^2$ at place, the minima of the potential is always negative.

\section{Circular strings in $(AdS_3)_\varkappa$}
\subsection{Solution of the string equations}
In this section we discuss circular string solutions in $(AdS_3)_\varkappa$ and talk about the various characteristics of it. The deformed $AdS$ is very unique in the regard that there is a radial singularity surface present in the background which appears in the scalar curvature too. So in general the $(AdS)_\varkappa$ has both positive and negative curvature regions depending on values of $\varkappa$. We will see how this singularity will affect the circular string solutions. Let us first start with the relevant metric, which we can get from (\ref{metric1}) by putting $\rho\to \cosh\rho$ and putting the sphere coordinates to be constants, 
\be
ds^2= -\frac{\cosh^2\rho}{1-\varkappa^2\sinh^2\rho}~dt^2+\frac{1}{1-\varkappa^2\sinh^2\rho}~d\rho^2+ \sinh^2\rho ~d\psi^2 \ .
\ee
We will use the standard circular string ansatz in this case
\be
t = t(\tau),~~~\rho = \rho(\tau),~~~\psi=m\sigma
\ee
The equation of motion for $\rho$ has the following form
\be 
\dot{\rho}^2 = \frac{\mathcal{E}^2}{\cosh^2\rho}(1-\varkappa^2\sinh^2\rho)^2- m^2\sinh^2\rho(1-\varkappa^2\sinh^2\rho) \ .
\ee
We can check that the equations of motion exactly matches with the Virasoro constraints for proper choice of integration constants.
Here the conserved charge $\mathcal{E}$ is defined as
\be
\mathcal{E}= \frac{E}{\sqrt{\hat{\lambda}}} = \frac{\cosh^2\rho}{1-\varkappa^2\sinh^2\rho}\dot{t} \ .
\ee
The above equation of motion for $\rho$ can be easily written in the form ,
\be
\dot{\rho}^2 = \frac{m^2(1-\varkappa^2\sinh^2\rho)(\sinh^2\rho_0^{+}-\sinh^2\rho)(\sinh^2\rho - \sinh^2\rho_0^{-})}{\cosh^2\rho} \ ,
\ee
where the roots are
\be
\sinh^2\rho_0^{\pm}= \frac{-(m^2+\mathcal{E}^2\varkappa^2) \pm \sqrt{(m^2+\mathcal{E}^2\varkappa^2)^2+4\mathcal{E}^2 m^2}   }{2m^2} \ .
\ee
We can easily see here that for the mentioned roots,
\be
\sinh^2\rho_0^{-}< 0;~~\sinh^2\rho_0^{+}> 0 \ ,
\ee
and the condition that $\sinh^2\rho_0^{\pm}\in \mathbb{R}$ is automatically satisfied. But there is another condition to be respected here for consistent string propagation, i.e.
\be
\dot{\rho}^2\geq 0
\ee
Due to this we can easily see that the string can only go up to a particular value of $\rho$ and would not be able to cross any further. This particular value is given by
\be
\rho  = \sinh^{-1}\frac{1}{\varkappa} \ ,
\ee
which is the same radial position where the so called `singularity surface' associated with the deformed metric is present. So it is quite clear that just like folded or other kinds of strings, the circular string will also reach upto this surface and no further. The way to avoid this problem has been to suggest that the singularity surface acts as an ad-hoc boundary for the deformed space and introduce the tortoise like coordinate transformation,
\cite{Kameyama:2014vma}
\be
\frac{\cosh\rho}{\sqrt{1-\varkappa^2\sinh^2\rho}}=\cosh\chi.
\ee
With this coordinate transformation we can actually map the coordinates $\rho\in [0, \sinh^{-1}\frac{1}{\varkappa})$ to $\chi\in[0,\infty)$. This is the same as saying that the unphysical region is simply not contained in the coordinate system used here. Now, the metric reads
\be
ds^2 = -\cosh^2\chi~dt^2+ \frac{1}{1+\varkappa^2\cosh^2\chi}~d\chi^2+ \frac{\sinh^2\chi}{1+\varkappa^2\cosh^2\chi}~d\psi^2
\ee
We can change the ansatz to include the new radial coordinate having the same functional dependence
\be
\chi = \chi(\tau).
\ee
The conserved charge can now be written as
\be
\mathcal{E} = \cosh^2\chi~\dot{t} \ .
\ee
In this changed coordinate system we can write the equation of motion for $\chi$ in the following form
\be
\dot{\chi}^2 = \frac{\mathcal{E}^2}{\cosh^2\chi}(1+\varkappa^2\cosh^2\chi)-m^2\sinh^2\chi
\ee
By putting $\sinh\chi =x$ we get
\be
\dot{x}^2 = m^2(x^2-R_-)(R_+-x^2)
\ee
Where the roots of the equation are
\be
R_\pm = \frac{(\mathcal{E}^2\varkappa^2-m^2)\pm\sqrt{(\mathcal{E}^2\varkappa^2-m^2)^2+4m^2\mathcal{E}^2(1+\varkappa^2)}}{2m^2} \ .
\ee
The solution for this string configuration has the form
\be
x = \sinh\chi(\tau) =  \sqrt{\frac{-R_- R_+}{R_+ -R_-}}~\textbf{sd}\left( m\sqrt{R_+ - R_-}\tau | \frac{R_+}{R_+ - R_-}\right).
\ee
The condition for having a oscillating solution is given by the periodicity condition on the Jacobi function, i.e.
\be
0 < \frac{R_+}{R_+ - R_-}<1.
\ee
This is also supplemented by the fact that both of the roots have to be real with $R_+ >0$ and $R_-<0$, which are automatically taken care of as we have discussed. The above condition then translates to
\be
(\mathcal{E}^2\varkappa^2-m^2)-\sqrt{(\mathcal{E}^2\varkappa^2-m^2)^2+4m^2\mathcal{E}^2(1+\varkappa^2)}<0,
\ee
Which is also taken care of by the condition $R_-<0$.
Now as usual, for various values of $\varkappa$ the oscillating nature of the solution will change. Since for obvious reasons we would be interested to find out the large $\varkappa$ characteristics of the solution. It can be shown in this limit, for any value of the energy $\mathcal{E}$,
\be
\frac{R_+}{R_+ - R_-} = 1 -\frac{m^2}{\mathcal{E}^2\varkappa^2}+ \mathcal{O}\left(   \frac{1}{\mathcal{E}^4\varkappa^4} \right).
\ee
So it is clear that in the limit $\varkappa\to\infty$ the modulus of the oscillatory solution actually proceeds to touch $1$ and the oscillatory behaviour of the string is lost in this limit. We can also argue this from a physical viewpoint as the large $\varkappa$ limit means that the `singularity surface' gradually coincides with the centre of the spacetime, i.e. $\chi = 0$ and the nature of the spacetime itself changes. 

\subsection{Semiclassical quantization and `long' string solution}
\begin{figure}

        \centering
        \begin{subfigure}[b]{0.52\textwidth}
                \includegraphics[width=\textwidth]{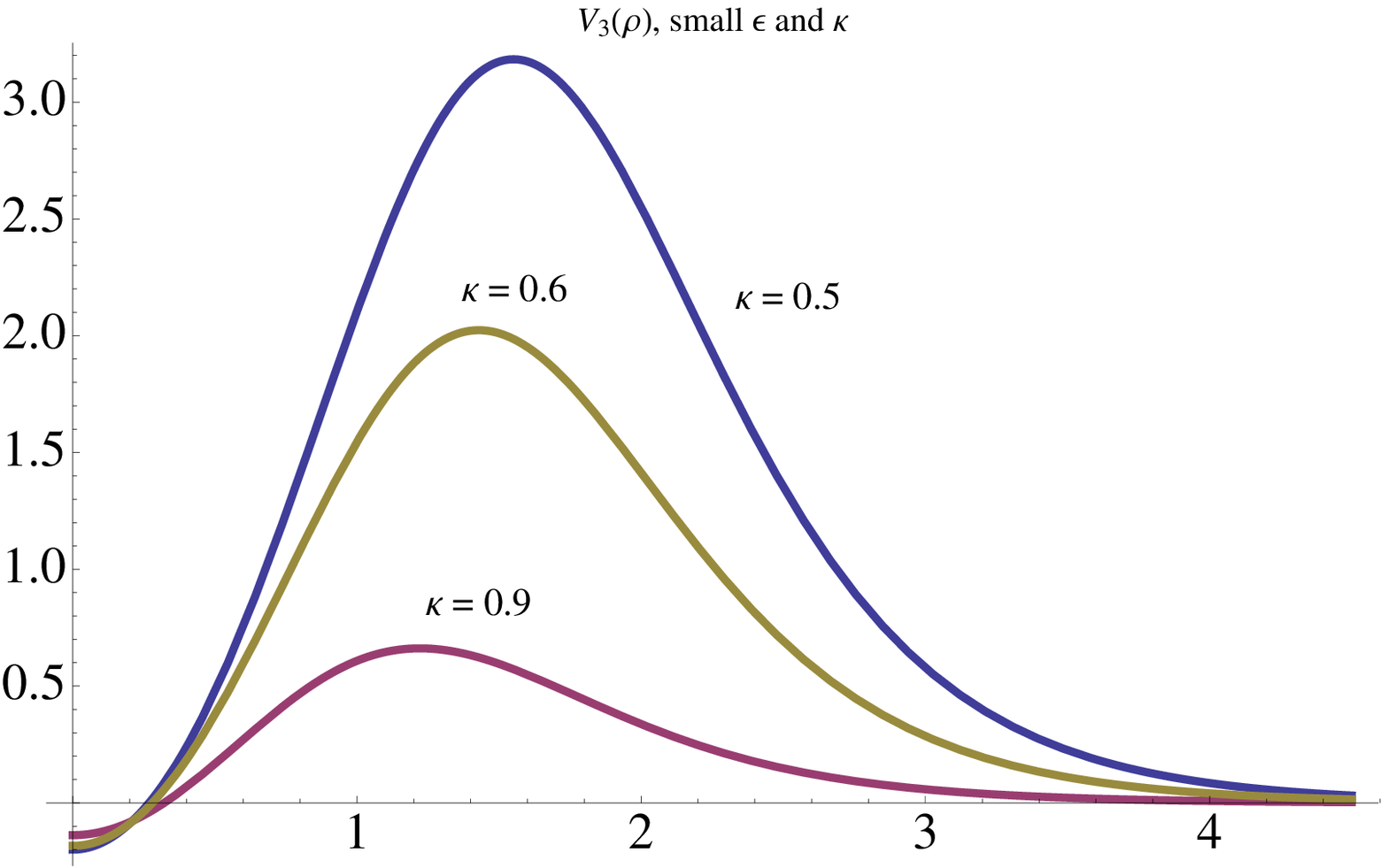}
                \caption{}
                \label{fig:1.1}
        \end{subfigure}%
        ~ %add desired spacing between images, e. g. ~, \quad, \qquad, \hfill etc.
        \begin{subfigure}[b]{0.52\textwidth}
                \includegraphics[width=\textwidth]{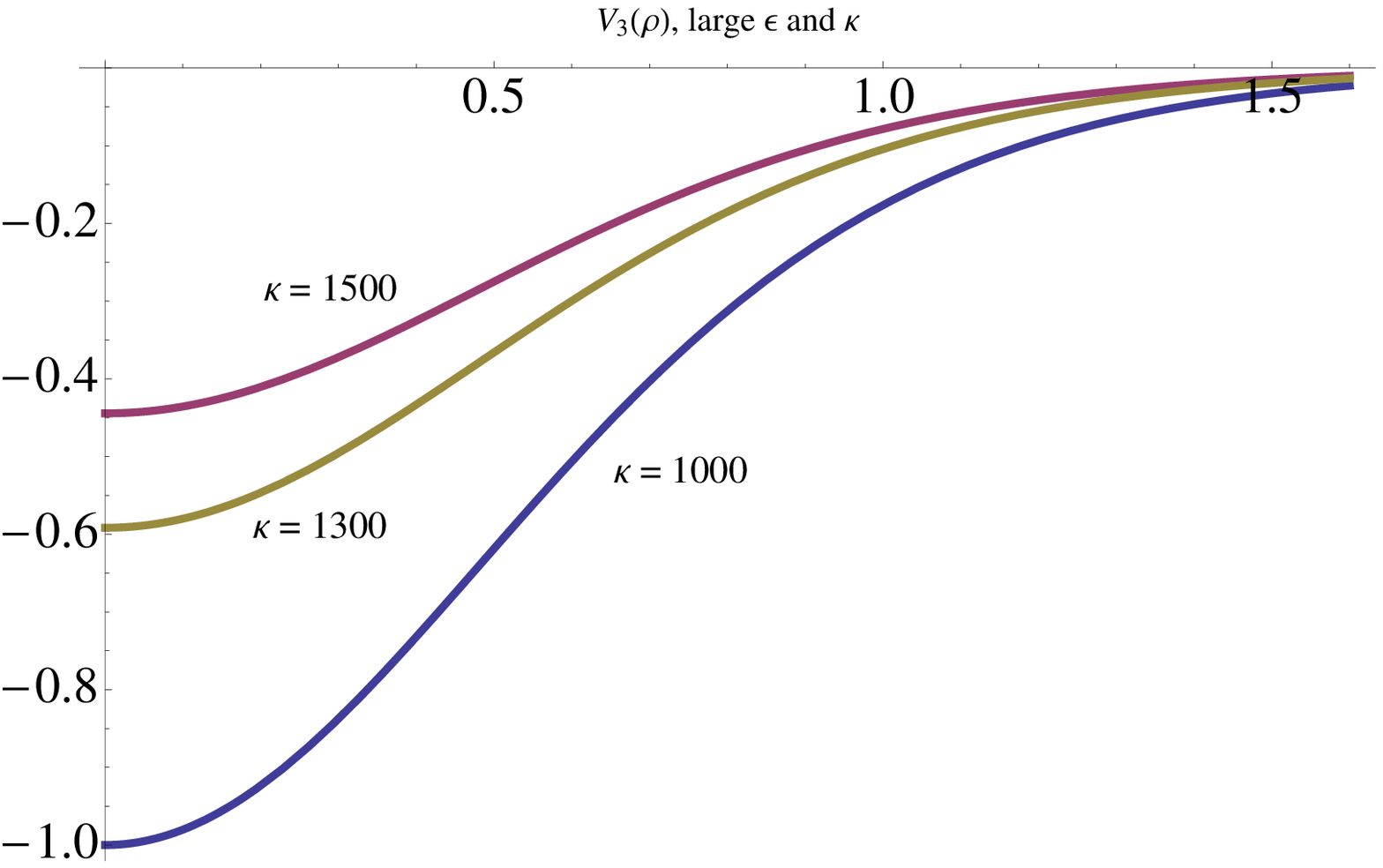}
                \caption{}
                \label{fig:1.2}
        \end{subfigure}

        \caption{The potential $V(\chi)$ has been plotted from (\ref{v3}) for different values of $\varkappa$ and $\mathcal{E}$. In (a) , the energy is small ($\mathcal{E}   = 0.5$) and $\varkappa$ is also small. In (b) both $\epsilon$ and $\varkappa$  are large ($\mathcal{E}   = 10^3$). One should notice the stark difference in the form of the potentials at extreme ends of the parameter space. For both cases we have taken $m=2$.}\label{fig:2}
\end{figure}

Let us now start with the quantization of the strings discussed in the last section. We note that the momenta associated to the coordinate $\chi$ has an expression
\be
\Pi_\chi = \frac{\dot{\chi}}{1+\varkappa^2 \cosh^2\chi} \ ,
\ee
so that the $\chi$ equation of motion can be written in the `particle in a potential' form
\be
\Pi_\chi^2 +V(\chi) = 0.
\ee
Where the effective potential has the form
\be
V(\chi) = -\frac{\mathcal{E}^2}{\cosh^2\chi(1+\varkappa^2 \cosh^2\chi)}+\frac{m^2\sinh^2 \chi} {(1+\varkappa^2 \cosh^2\chi)^2}. \label{v3}
\ee
Since the $AdS$ circular string can have both small and large energy limits the characteristics of this potential are bound to be different in different limits. The position of the maxima for this potential also varies depending on the value of $\varkappa$. We plot the potential for different values of 
the parameters in figure (\ref{fig:2}). 

Now, with a bit of algebra, we write the oscillation number in the form
\be
\mathcal{N}  = \frac{2}{\pi}\int_0^{\chi_{m}}\frac{d\chi}{1+\varkappa^2 \cosh^2\chi}\sqrt{\frac{\mathcal{E}^2}{\cosh^2\chi}(1+\varkappa^2\cosh^2\chi)-m^2\sinh^2\chi
} \ .
\ee
We must remember here that $\mathcal{N}$ can be thought of as the deformed version of the total oscillator number, which has to be even for the closed string sector. We will take care of the integral for the oscillation number as we have done earlier, i.e. by taking the derivative w.r.t. $m$ and putting $\sinh\chi = z$, which leads to the following integral,
\be
\frac{\partial \mathcal{N}}{\partial m}  = -\frac{2}{\pi}\int_0^{\sqrt{R_+}}\frac{z^2}{(1+\varkappa^2+\varkappa^2 z^2)\sqrt{(R_+ - z^2)(z^2 - R_-)}} \ ,
\ee
where the roots have been presented as in the last section. This integral results in the combination of elliptic integrals  of the form
\be
\frac{\partial \mathcal{N}}{\partial m}  = -\frac{2}{\pi}\frac{1}{\sqrt{-R_-}~\varkappa^2} \bigg[ \mathbf{K} \left( \frac{R_+}{R_-}\right)  - \mathbf{\Pi} \left( -\frac{R_+ \varkappa^2}{1+\varkappa^2} , \frac{R_+}{R_-}  \right) \bigg].
\ee
The above expression can be expanded in the limit of large $\mathcal{E}$ (and finite $\varkappa$) and integrated over $m$ to get the expression for the oscillation number in terms of other parameters. In this case it has the most unusual form,
\be
\mathcal{N} = \mathcal{N}_0(\mathcal{E},\varkappa)+ \frac{m^2}{\pi\varkappa^3\mathcal{E}}  \log \left[ \frac{a_1 m}{\varkappa \mathcal{E}}   \right] + \mathcal{O}\left(  \frac{1}{\mathcal{E}^3} \right),
\ee

Which is a completely new scaling for such long strings and does not reduce to the usual form of oscillation number for $AdS$ strings. The constant $a_1 =0.1516 $. The $\mathcal{N}_0$ is the integration constant here which is nothing but the $m = 0$ integral of $\mathcal{N}$. This limit only just corresponds to a massless geodesic in this background which reaches from $\chi = 0$ to $\chi = \infty$. The integral has a form,
\be
\mathcal{N}_0 = \frac{2\mathcal{E}}{\pi}\int_0^{\infty} \frac{d\chi}{\cosh\chi \sqrt{1+\varkappa^2\cosh^2\chi}} = \frac{2\mathcal{E}}{\pi} \tan^{-1}\frac{1}{\varkappa}.
\ee
It is worth noticing that the usual leading term for circular long strings in undeformed $AdS$ can actually be recovered by putting $\varkappa = 0$. However the scaling behaviour like $\frac{1}{\mathcal{E}}  \log \left[ \frac{1}{\mathcal{E}}   \right]$ remains very mysterious.

\section{Conclusion}
In this short note, we have discussed various circular string configurations in the one parameter (or $\varkappa$) deformed $AdS_3\times S^3$ background. The background is unique and very interesting by itself, and as expected the string solutions from the string sigma model action gives various insights into the structure of the background. We have discussed various string solutions in the two-spheres belonging to $S^3$. After finding the exact solutions in terms of Jacobi functions we have proven that the solutions would obey the same discrete symmetries of the total background. We also have discussed about the same string solutions in the $AdS$ part. A remarkable speciality of this background is a presence of a singularity surface where curvature invariants blow up. To keep our spacetime timelike, we have performed a coordinate transformation and discussed strings that go up to this surface, which might act like an ad-hoc boundary. But since the deformation is independent of the coordinates, we  arrive at a energy expansion for a novel `long' strings, which is new and does not reduce to the undeformed result even if we take $\varkappa\to 0$. The explanation of such kind of leading order behaviour is not entirely clear to us. The nature and validity of this special scaling remains to be checked in detail. The other issue that one can perform from here is to use the deformed Neumann-Rosochatius integrable system for the deformed setup as described in \cite{Arutyunov:2014cda}, and try to find out more general circular and elliptic solutions following the algorithm. But due to very complicated nature of the integrals of motion, this appears to be a daunting task. We hope to address this question in a future publication.  

\section*{Acknowledgements} 
AB would like to thank Saha Institute of Nuclear Physics, Kolkata for kind hospitality, during which most of this work was done. KLP would like to thank DESY theory group for hospitality under SFB fellowship where a part of this work was done.

\end{document}